\documentclass[conference,a4paper,final]{IEEEtran}
\usepackage[latin1]{inputenc}
\usepackage[T1]{fontenc}
\usepackage{textcomp}
\usepackage{xspace}
\usepackage{ifpdf}

\usepackage{cite}

\ifCLASSINFOpdf
  \usepackage[pdftex]{graphicx}
  \usepackage[pdftex,hypertexnames=false,breaklinks=true]{hyperref}
  \DeclareGraphicsExtensions{.pdf,.jpeg,.png}
\else
  \usepackage[dvips]{graphicx}
  \DeclareGraphicsExtensions{.eps}
\fi
\usepackage[cmex10]{amsmath}
\usepackage{algorithm}
\usepackage{algpseudocode}

\usepackage{mdwmath}
\usepackage{mdwtab}
\usepackage[caption=false,font=footnotesize]{subfig}
\usepackage{fixltx2e}

\usepackage{stfloats}

\usepackage{url}

\usepackage{amssymb}
\usepackage{color}
\usepackage{tikz}
\usetikzlibrary{matrix,arrows}
\tikzset{level distance = 4ex, sibling distance = 4ex, every node/.style={fill=gray!30,circle,inner sep=0.5pt}, baseline=(current bounding box.west)}
\newcommand{\DEF}{
  \mathbin{\smash[t]{\overset{\scriptscriptstyle\mathrm{def}}{=}}}}
\newcommand{\ex}{\ensuremath{\mathbin{\diamond}}\xspace}
\newcommand{\nil}{\ensuremath{\mathord{nil}}\xspace}
\newcommand{\TExt}{\ensuremath{\operatorname{\texttt{TPM\_Tree\_Extend}}}\xspace}
\newcommand{\vload}{\ensuremath{\operatorname{\texttt{TPM\_Reduced\_Tree\_Verify\_Load}}}\xspace}
\newcommand{\vred}{\ensuremath{\operatorname{\texttt{TPM\_Reduced\_Tree\_Verify}}}\xspace}
\newcommand{\vfull}{\ensuremath{\operatorname{\texttt{TPM\_Tree\_Node\_Verify}}}\xspace}
\newcommand{\vupd}{\ensuremath{\operatorname{\texttt{TPM\_Reduced\_Tree\_Update}}}\xspace}
\newcommand{\vvupd}{\ensuremath{\operatorname{\texttt{TPM\_Tree\_Node\_Verified\_Update}}}\xspace}
\newcommand{\TQ}{\ensuremath{\operatorname{\texttt{TPM\_Tree\_Node\_Quote}}}\xspace}
\newcommand{\binr}[1]{
            \ensuremath{\mathord{\left\langle#1\right\rangle}}\xspace}
\newcommand{\ChiralEx}[3]{
            \ensuremath{\mathord{\left\langle\vphantom{#2 #3}#1\right|\! \mathbin{#2} \!\left| #3\vphantom{#2 #1}\right\rangle}}\xspace}
\begin{document}
%
\title{Secure Operations on Tree-Formed Verification Data}
\author{\IEEEauthorblockN{Andreas U. Schmidt and Andreas Leicher}
\IEEEauthorblockA{Novalyst IT AG\\
Robert-Bosch-Straße 38,
61184 Karben, Germany\\
Email at: http://andreas.schmidt.novalyst.de/}
 \and
 \IEEEauthorblockN{Yogendra Shah and Inhyok Cha}
 \IEEEauthorblockA{InterDigital Communications, LLC\\
 781 Third Avenue, King of Prussia, PA 19406\\
 Email: \{yogendra.shah,inhyok.cha\}@InterDigital.com}}


%


\maketitle

\begin{abstract}
%
%
We define secure operations with tree-formed, protected  verification data registers.
Functionality is conceptually added to Trusted Platform Modules (TPMs) to handle
Platform Configuration Registers (PCRs) which represent roots of hash trees protecting the 
integrity of tree-formed Stored Measurement Logs (SMLs).
This enables verification and update of an inner node of an SML and even attestation to
its value with the same security level as for ordinary PCRs.
As an important application, it is shown how certification of SML subtrees 
enables attestation of platform properties.
\end{abstract}


%
\IEEEpeerreviewmaketitle
\section{Introduction}
%
%
The process of building trust in computing platforms follows
a unique, common pattern~\cite{AUS09B}. All components of the platform
are measured by a protected entity on the platform before they are loaded and executed.
The generation of a chain of trust, which extends without gaps from system boot up 
to the current system state is an important concept for a Trusted Computing System.
To prevent unmonitored execution of code between measurement 
and actual execution, every component is required to measure 
and report the following component before executing it, while 
this measurement process is protected by the root of trust for measurement.
\textit{Verification data} is compiled from the measurement values by a protected
operation and stored in protected storage. The verification data identifies, after
completion of secure start up, the platform's state uniquely.
Specified by the Trusted Computing Group (TCG), the most important 
embodiments of these processes are \textit{authenticated boot}~\cite{PCCLIENTBIOS}
and \textit{secure boot}~\cite{MTMREFARC}.
Secure boot includes a local verification and enforcement engine that lets 
components start only if their measurements are equal to trusted reference values.

In~\cite{tree01}, a modification of the extend operation of TPMs was described which
allows a verification data register, i.e., a PCR, 
to protect the root of a Merkle hash tree, which is stored in a \textit{tree-formed SML}.
This exposes a new TPM command, \TExt, which assures that a sequence of measurements
of system components and/or data are organised into a binary tree of which a designated
verification data register is the root.
Tree-formed SML extends the verification data from the root register to a complex data
structure. It has various usages, for instance the efficient search for \textit{failed}
components, i.e., leaf measurements with undesired values.

In the present paper, we add more trusted functionalities to operate on tree-formed
verification data. We show how inner nodes of a tree-formed SML with its
root protected in a verification data register, can be verified for integrity, and updated with
a new value, in a controlled way maintaining the overall security level.
Finally, we introduce a variant of the $\texttt{TPM\_Quote}$ command for inner tree nodes,
which attests to their integrity precisely as $\texttt{TPM\_Quote}$ does for an ordinary PCR's value.
With the defined set of commands, the integrity measurement functionality
of a TPM is complemented by a comprehensive set of commands operating with tree-formed PCRs and SMLs. 
Using them, tree-formed verification and validation data can be used with far more flexibility and
expressiveness than linearly chained TPM PCRs and SMLs.

Section~\ref{sec:preparations} introduces a basic system model and notation.
Section~\ref{sec:tree-update} defines the TPM command extensions described above, and also
proposes some pertinent structural extensions and basic usage categories for such operations.
As a central use case for the introduced commands, Section~\ref{sec:subtr-cert} exhibits a 
protocol for certification of the root node
of a subtree in a tree-formed SML by a trusted third party.
We conclude with Section~\ref{sec:conclusion} with a brief assessment of the tree-formed
platform validation approach and an outlook to future work.
The reader is referred to~\cite{tree01} for an overview of related work. More related work
is discussed as the concepts are developed.
\section{Preparations}\label{sec:preparations}
In this section, the minimal necessary elements and capabilities of a platform, which are required
subsequently, are described. While we are leaning on TCG-nomenclature and some concepts, it will
be clear from these minimal requirements, that the concepts developed in the following
sections are not restricted to platforms and secure hardware elements adhering to TCG
standards, e.g., TPMs~\cite{TPMMAIN} and systems designed according to the PC Client specification~\cite{PCCLIENTBIOS}.
\subsection{System Model}\label{sec:system-model}
The tree-formation variant of the extend operation defined in~\cite{tree01}, operates
inside a TPM, takes only a single measurement value as input, and is otherwise inert
with regard to the system outside the TPM.
This is not the case for the update function which is introduced in the present paper.
The latter operates on a certain number $r$ of verification data registers $V$,
$\mathcal{V}\DEF\{V_1,\ldots,V_r\}$ 
protected inside a TPM, and on the hash tree data
stored outside the TPM in less protected storage. 
That is, the hash tree contained in the Stored Measurement Log (SML) is managed
by a Trusted Software Stack (TSS) which is authorised to access the TPM functions
necessary for the update operations. TSS calls TPM via an authorised, integrity-protected
command interface. Note that, while we use TCG parlance for practical reasons,
the concepts presented here and in~\cite{tree01} are not restricted to a TCG TPM
and platform. We only assume a hardware-protected set of verification data registers and
an extend operation. The latter is defined by the ordinary TPM extend operation
\begin{equation}
	 V \gets  V\diamond m \DEF H \left( V \|  m  \right),
  \label{eq:extend}
\end{equation}
where $V$ denotes a verification data register, $H$ is a collision-resistant
hash function (SHA-1 in case of the TPM), and $m=H(\text{data})$ is a measurement value.
In the following \ex is used liberally with arbitrary registers $V$ as arguments, where
no confusion can arise. 
\subsection{Conventions}\label{sec:reduced-trees}
We assume that the SML contains a binary tree of depth $d$ resulting from a binary 
one-way operation, such as the Merkle hash tree~\cite{Merkle1989,10.1109/SP.1980.10006} 
produced by the \TExt command
introduced in~\cite{tree01}. Natural coordinates for 
inner nodes and leaves are binary strings of length $1,\ldots,d$, where the length
$\ell$ of the string is the level in the tree on which the node resides.
Let $\mathbf{n}$ be an inner node or leaf and write 
$\mathbf{n}\sim\binr{\mathbf{n}}=(n_1,\ldots,n_\ell)\in\{0,1\}^{×\ell}$ for the binary 
representation of the coordinates of $\mathbf{n}$. 
Let $\binr{\mathbf{n}}_k=n_k$, $k=1,\ldots,\ell$, be the 
$k$-th digit of \binr{\mathbf{n}}.
Where no confusion can arise, we identify a node with its value (e.g., 160-Bit hash value) 
in the SML, while distinguishing it from its coordinate.
Otherwise we write $\mathbf{n}=(n,\binr{\mathbf{n}})$ for the value-coordinate
pair of a node.

The \textbf{trace} $\mathbf{T}$ of $\mathbf{n}$ is the ordered list of all inner nodes on the path
from $\mathbf{n}$ to the root, including $\mathbf{n}$, i.e.,
\begin{equation}
  \label{eq:trace}
  \mathbf{T}(\mathbf{n})=(\mathbf{t}_1,\ldots,\mathbf{t}_\ell),\ \text{where }
 \mathbf{t}_k\sim(n_1,\ldots,n_k).
\end{equation}
The natural partial order of nodes is written as $\mathbf{m}\leq \mathbf{n}$,
which is equivalent to $\mathbf{n}\in \mathbf{T}(\mathbf{m})$.
The partial order extends to sets $M$, $N$ of nodes by
setting $M\leq N$ iff $ \forall \mathbf{m}\in M\colon\exists \mathbf{n}\in N\colon \mathbf{m}\leq \mathbf{n}$.

The \textbf{reduced tree} $\mathbf{R}$ of $\mathbf{n}$ is the list of all siblings of its trace.
This is readily expressed in natural coordinates.
\begin{equation}
  \label{eq:reduced}
  \mathbf{R}(\mathbf{n})=(\mathbf{r}_1,\ldots,\mathbf{r}_\ell),\ \text{where }
 \mathbf{r}_k\sim(n_1,\ldots,¬ n_k),
\end{equation}
where $¬$ denotes binary negation.

We use the hash chain operation $x\ex y\DEF H(x\| y)$, with fixed-length input hash values $x$, $y$, 
in a variant which makes argument order dependent on a binary parameter. 
We set, for $c\in\{0,1\}$,
\[
 \ChiralEx{x}{c}{y}=
\begin{cases}
 x \ex y &\text{if $c=1$;}\\
 y \ex x &\text{if $c=0$.}
\end{cases}
\]
This \textit{chiral Merkle-Damg\r{a}rd operation} is a
version of an extend operation which allows to distinguish between left and right siblings
in a tree and calculate their parent node in the correct order. Neglecting implementation
issues, we assume that the (extended) TPM is capable of performing the operation \ChiralEx{\cdot}{\cdot}{\cdot}
internally.

In many cases, the hash tree stored in the SML may be incomplete, i.e., contain
empty leaves and inner nodes, denoted by \nil.
For the consistent treatment of \nil nodes in a Merkle hash tree, 
it is useful to assume that \nil is a two-sided unit for the operation \ex, i.e.,
\begin{equation}
  \label{eq:exunit}
  x\ex\nil = \nil\ex x = x, \text{and } \nil\ex\nil=\nil.
\end{equation}
This is a 
re-interpretation of the usual TPM extend operation and can also be used to
model a direct write to a $V\in\mathcal{V}$, by first resetting $V$ to \nil and then
performing $V\ex x$ for some value $x$.
For the implementation of this convention, 
we may assume that \nil is be represented
as a flag of verification data registers  
and the inputs and output of \ChiralEx{\cdot}{\cdot}{\cdot}.
For a $V$, the \nil flag may be set by a particular reset command.
When \nil is encountered as the input
of an extend operation to a $V$, then logic of the TSS, or a TPM modification, may
prevent execution of the extend and write to the PCR directly. 
\section{Secure Operations with Tree Nodes}\label{sec:tree-update}
This main section presents the operational extensions of a standard TPM to operate
securely with tree-formed SMLs. The protection goal is to achieve the same assurance level for
inner nodes and leaves of such an SML, as for a conventional verification data register
value, protected in a PCR. We first describe the update of a root by a new node value,
and then show further structural and command extensions for use with tree-formed 
verification data.

The strategy for a secure update of an inner node or leaf of a SML tree is as follows.
First, the current value of that node needs to be verified for authenticity.
This is done by recalculating the root of the tree, protected in a register $V$,
(which is kept fixed in the remainder of the paper to simplify presentation)
using the data contained in the reduced hash tree associated with the node.
This verification must be a protected operation inside the TPM, called \vload.
It also loads the verified reduced tree data into a set of verification data registers
for use with the subsequent update operation \vupd.
This function takes a new value for the node to be updated, and uses the reduced tree
data to update the parent nodes up to the root $V$. Both commands may be used 
separately for various purposes, e.g. standalone node integrity verification.
For convenience, they may also be combined into a single node and root update
command.
\subsection{Verified Load of a Reduced Tree}\label{sec:ver-load}
Suppose $\mathbf{n}$ is the node of an SML tree of depth $d$ at level $\ell\leq d$
with root protected in a verification data register $V\in\mathcal{V}$.
The first step to update $V$ with a new value for $\mathbf{n}$, 
is to verify that the reduced tree $\mathbf{R}(\mathbf{n})$ is untampered in the SML.
To maintain the security level of $V$, this verification
needs to be performed by a TPM-protected operation as well. For this, TSS calls
\vload with arguments $(\binr{\mathbf{n}},\mathbf{n},\mathbf{R}(\mathbf{n}))$.
Choose $\ell+1$ available registers from $\mathcal{V}$ and call them $B_1,\ldots,B_\ell$,
and $V^\ast$. Algorithm~\ref{alg:vload} shows how an SML node is verified and its reduced
tree is loaded into a set of verification data registers.

The chiral extend used centrally in this algorithm ensures correct order of the child nodes in 
the calculation of their parent element on the trace of $\mathbf{n}$.
The TSS obtains the calculated trace $\mathbf{T}(\mathbf{n})$ and the verification status as
return values. Algorithm~\ref{alg:vload}
requires $\ell+1$ additional verification data registers. 
\begin{algorithm}[h]
\begin{algorithmic}[1]
\Require $B_1,\ldots,B_\ell, V^\ast\in\mathcal{V}, (\binr{\mathbf{n}},\mathbf{n},\mathbf{R}(\mathbf{n}))$
\Ensure $B_1 \gets \mathbf{r}_1,\ldots,B_{\ell} \gets \mathbf{r}_\ell, V^\ast \gets \mathbf{n}$ \hspace{\fill}
\linebreak \Comment Initialise buffer with reduced tree and node to verify.
\For{$k=\ell,\ldots, 1$}
  \State $V^\ast\to\text{TSS}$ 
  \State $V^\ast \gets \ChiralEx{B_k}{\binr{\mathbf{n}}_k}{V^\ast}$
\EndFor
\If{$V\equiv V^\ast$}
  \State \textbf{return} $V$
\Else
  \State \textbf{return} ``verification error''
\EndIf
\end{algorithmic}
\caption{\vload}
\label{alg:vload}
\end{algorithm}

A simple variant of Algorithm~\ref{alg:vload} can operate using only a single
verification data register, by processing the reduced tree sequentially, without storing
the reduced tree inside the TPM. This auxiliary command $\texttt{TPM\_Reduced\_Tree\_Verify}$
may be useful for a plain verification of the
SML by the TSS or another party. This is shown in Algorithm~\ref{alg:verify}.
The serialisation of $\mathbf{R}(\mathbf{n})$ required by this variant may be done 
using an input buffer realised in a software layer below the TSS, e.g., a TPM device
driver, or by corresponding TPM internal logic. 

\begin{algorithm}[h]
\caption{\vred\label{alg:verify}}
\begin{algorithmic}[1]
\Require $B\in\mathcal{V}, (\binr{\mathbf{n}},\mathbf{n},\mathbf{R}(\mathbf{n}))$
\Ensure $B \gets \mathbf{n}$ \Comment Initialise buffer with node to verify.
\For{$k=\ell,\ldots, 1$}
  \State $B\to \text{TSS}$
  \State $B \gets \ChiralEx{\mathbf{r}_k}{\binr{\mathbf{n}}_k}{B}$
\EndFor
\If{$V\equiv B$}
  \State \textbf{return} $V$
\Else
  \State \textbf{return} ``verification error''
\EndIf
\end{algorithmic}
\end{algorithm}

Like the original tree formation algorithm of~\cite{tree01}, Algorithms~\ref{alg:vload}
and~\ref{alg:verify} use non-standard operations, in particular chiral extend. 
Since the output target of chiral extend is always a verification data register,
the operation can be implemented by loading the other argument into another verification
data register (if it is not already there, as in Algorithm~\ref{alg:vload}), and
preceding the TPM-internal operation \ex with a register swap, depending on the
middle argument of chiral extend. This ensures the same protection level for
all arguments.
\subsection{Full Node Verification}
The verification performed by algorithms~\ref{alg:vload} and~\ref{alg:verify}
has a limited meaning since it only assures the integrity of the input node value
with respect to the input reduced tree. In case of an integrity breach of the 
SML tree, more detailed information is desirable. We can obtain at least the tree
level at which an integrity breach occurs, by performing the validation strategy
via downward tree-traversal described in~\cite{tree01}.

The command \vfull shown in Algorithm~\ref{alg:fullverify} 
returns the level at which an incorrect reduced tree and/or trace element first broke 
the integrity chain from the root to $\mathbf{n}$. It obviously does \textit{not} allow 
to determine which sibling broke the chain. Further diagnostics would only be possible when
a reference tree is available, see~\cite{tree01}.
\begin{algorithm}[th]
\caption{\vfull\label{alg:fullverify}}
\begin{algorithmic}[1]
\Require $B,C,D\in\mathcal{V}, (\binr{\mathbf{n}},\mathbf{R}(\mathbf{n}),\mathbf{T}(\mathbf{n}))$
\Ensure $C \gets  V$ \Comment Initialise comparison register with root.
\For{$k=1,\ldots, \ell$}
  \State $B,D \gets \mathbf{t}_k$ \Comment Load trace child into buffers 
  \State $B \gets \ChiralEx{\mathbf{r}_k}{\binr{\mathbf{n}}_k}{B}$
  \If{$C \equiv B$} \hspace{\fill}
\linebreak \Comment Compare result with current parent, and if OK,
    \State $C\gets D$ \hspace{\fill}
\linebreak \Comment make the trace element just verified the new parent.
  \Else
    \State \textbf{return} ``verification error at level '' $||\  k, \ B$
  \EndIf
\EndFor
\State \textbf{return} ``OK''
\end{algorithmic}
\end{algorithm}
  
\subsection{Root Update}\label{sec:root-update}
Assume that \vload has been performed for a node $\mathbf{n}$ which shall now be updated 
with a new value $\mathbf{n}'$. 
\begin{algorithm}[ht]
\caption{\vupd\label{alg:vupd}}
\begin{algorithmic}[1]
\Require $\binr{\mathbf{n}},\ \mathbf{n}'$
\Ensure $B_1=\mathbf{r}_1,\ldots,B_\ell=\mathbf{r}_\ell$
\State $V \gets \mathbf{n}'$
\For{$k=\ell,\ldots, 1$}
  \State $V \to \text{TSS}$
  \State $V \gets \ChiralEx{B_k}{\binr{\mathbf{n}}_k}{V}$
\EndFor
\State \textbf{return} $V$
\end{algorithmic}
\end{algorithm}The command \vupd is called with argument $\mathbf{n}'$ and may exclusively operate
on the result of a determined, preceding \vload, which also fixes the node coordinate
$\binr{\mathbf{n}}$ and the register $V$ to be updated. To achieve this binding in a 
command sequence, various methods can be employed. First, the TPM can store and manage 
states and additional data for tree operations as described in Section~\ref{sec:verif-data-regist}.
Furthermore the sequence of commands \vload and \vupd should be bound cryptographically, for
instance by rolling nonces as implemented by TPM protected OIAP/OSAP authorised 
command sessions~\cite[p. 60ff]{TPMMAIN}.
Finally, the two commands may be joined to a single update command
\vvupd, with arguments $(\binr{\mathbf{n}},\mathbf{n}, \mathbf{n}', \mathbf{R}(\mathbf{n}))$.
The node update commands return the new trace of $\mathbf{n}'$ and the new value of $V$,
with which the TSS then updates the SML.
\subsection{Verification Data Register States}\label{sec:verif-data-regist-stat}
With the association of verification data registers to certain nodes or roots of hash trees,
and the associated commands \TExt (defined in~\cite{tree01}), \vload, \vupd, 
these registers $V$ acquire statefulness. States of particular importance may be
\begin{itemize}
\item\textbf{Active Root (AR)} signifying a root of an SML tree currently under construction by
the \TExt operation.
\item\textbf{Complete Root (CR)} signifying the root of a tree which is the completed result
of the measurement of a number of components, i.e., \TExt operations.
AR can transition to CR when the tree is full, i.e., contains 
$2^d$ leaf measurements, or triggered by the TSS if it is desired to close a tree at a certain
stage. A $V$ in CR state should be protected against further updates with \TExt, but may be accessed
by \vupd or even the normal $\texttt{TPM\_Extend}$ operation depending on policies and authorisation.
\item\textbf{Tree Build (TB)} signifying a register used to build an active tree in another, AR  register
by the \TExt operation.
\item\textbf{Reduced Tree Node (RT)} signifying the result of \vload, i.e., one of the
registers $B_k$. An RT $V$ must be protected 
until the corresponding \vupd, or another, authorised command occurs.
\end{itemize}
When more than one tree is managed, $V$s' states need to be associated to their respective
trees, e.g., using Unique Identifiers (UIDs). Furthermore node coordinates may need to be stored
for each or some register(s). 
These data could be held in a Verification Data Allocation Table (VDAT) inside the TPM, 
and managed by a Tree Data Management Unit (TDMU). 
\subsection{Quoting a Tree Node}\label{sec:verif-data-regist}
TPM protected verification of a node value enables a new core semantic for platform
validation by attestation. In particular, we can define a variant of $\texttt{TPM\_Quote}$
that attests to a certain node value. 
In its most elementary form, such a command \TQ is called with the same arguments
as $\texttt{TPM\_Quote}$ plus the arguments of \vred. It then executes Algorithm~\ref{alg:verify},
but additionally keeps a copy of $\mathbf{n}$ in another PCR $V'$.
Upon success it executes $\texttt{TPM\_Quote}$ on $V'$.
The receiver of such a quote should be made aware that the signature obtained is over an
SML tree's inner node. One possibility would be to change the fixed string contained
in the signed blob of the $\texttt{TPM\_Quote}$ command, which normally is 
``QUOT''~\cite[Part 3, line 2794 on page 161]{TPMMAIN}, to, say,
``TREEQUOT''.


Attestation to a node value with this command provides to the node the same security
assertion as quoting a verification data register (PCR) value with $\texttt{TPM\_Quote}$.
However, it bears the additional semantics that the value corresponds to some inner node
of an SML tree, i.e., it effectively attests to the state of a certain subtree of which
$\mathbf{n}$ is the root. To explicitly convey this semantics to a validator, additional data
may be included in the AIK (Attestation Identity Key) 
signed attestation package, e.g., a string ``$\texttt{Tree Node}$''.
The meaning of such an attribute can be sensibly strengthened, if it is only assigned
by \TQ if the root register is a controlled SML root register resulting from \TExt commands, i.e., it is 
in the CR state, cf.\ Section~\ref{sec:verif-data-regist-stat}.
This control should be part of the quote generation.

For the validation of an attestation message, the validator needs only the value
$n$ of the quoted node $\mathbf{n}=(n,\binr{\mathbf{n}})$.
More information transfer to the validator is in principle not necessary, therefore
the above description of \TQ follows a principle of minimal revelation.
A variant of the command may also sign the node coordinate \binr{\mathbf{n}},
if the position of the node in the SML tree matters for validation.
Extended validation data transferred to a validator could also include the
reduced tree of $\mathbf{n}$ and root verification data register, where this makes sense.

As a straightforward alternative, it would be possible to task the validator with the verification
of the reduced tree. This approach is used in~\cite{DBLP:conf/acsac/MoyerBSMJ09} to attest to the
integrity of Web pages delivered by a Web server, and to bind this to an attestation of the
server's state using the ordinary $\texttt{TPM\_Quote}$ command. This brings us to a variant realisation
of \TQ, simply as follows. The command receives as arguments the node value $n$, the node values of 
$\mathbf{R}(\mathbf{n})$, and a selector for the root $V$. The TPM signs\footnote{Note
that in~\cite{DBLP:conf/acsac/MoyerBSMJ09}, a workaround method is used to
bind a reduced tree to a quote from a TPM, by inserting a hash of these additional data
into the \textit{nonce} input of the $\texttt{TPM\_Quote}$ command, which is normally
used to guarantee freshness of the quote.} this (concatenated) data 
after controlling the CR state of $V$ and with a ``REDTREEQUOT'' fixed string attribute.

The first and second realisation variant for quoting an inner node represent
opposite possibilities, in the sense that the first puts verification load with the platform, while
the second puts it with the validator. Therefore, both may have different domains of practical
efficiency. For instance, many, distributed validating platforms as in M2M communication for the first,
and many, distributed validators (such as the Web clients in~\cite{DBLP:conf/acsac/MoyerBSMJ09}) for
the second variant. But note that the second variant has, by principle drawbacks with regard
to information revelation of the platform, since the validator is shown the complete state represented
by $V$. This may be detrimental to privacy.
\subsection{Baseline Applications}\label{sec:applications}
The various extensions of the TPM integrity measurement functionalities 
introduced in this paper and~\cite{tree01} can be grouped into the following
categories. 
\TExt is used in the (continuous) measurement process that builds a particular
SML tree with PCR-protected root. 
\vload, \vred, and ultimately \vfull are commands for
platform-internal diagnostics. Apart from the usage of \vload as a preparation step to
\vupd, they may be used to verify certain properties of a platform, represented by
SML subtrees, before other events can happen.
\vupd and \vvupd are used for the controlled update of subtrees.
Particular usages of an inner node update operation are
\begin{itemize}
  \item Update of a single system component. In this case the new value updates a leaf.
  \item Update of a system module represented by a subtree. In this case the root of the
new subtree updates an inner node of the original tree.
\end{itemize}
Finally, \TQ is the command which makes tree-formed SML usable for validation of a platform
by a remote party. It exhibits the key new element of validation using tree-formed data, 
namely the possibility to attest to subtrees representing only a defined part of the system
state. A particular use case is described in the next section.
\begin{figure*}[!t]
\centerline{\includegraphics[width=0.54\textwidth]{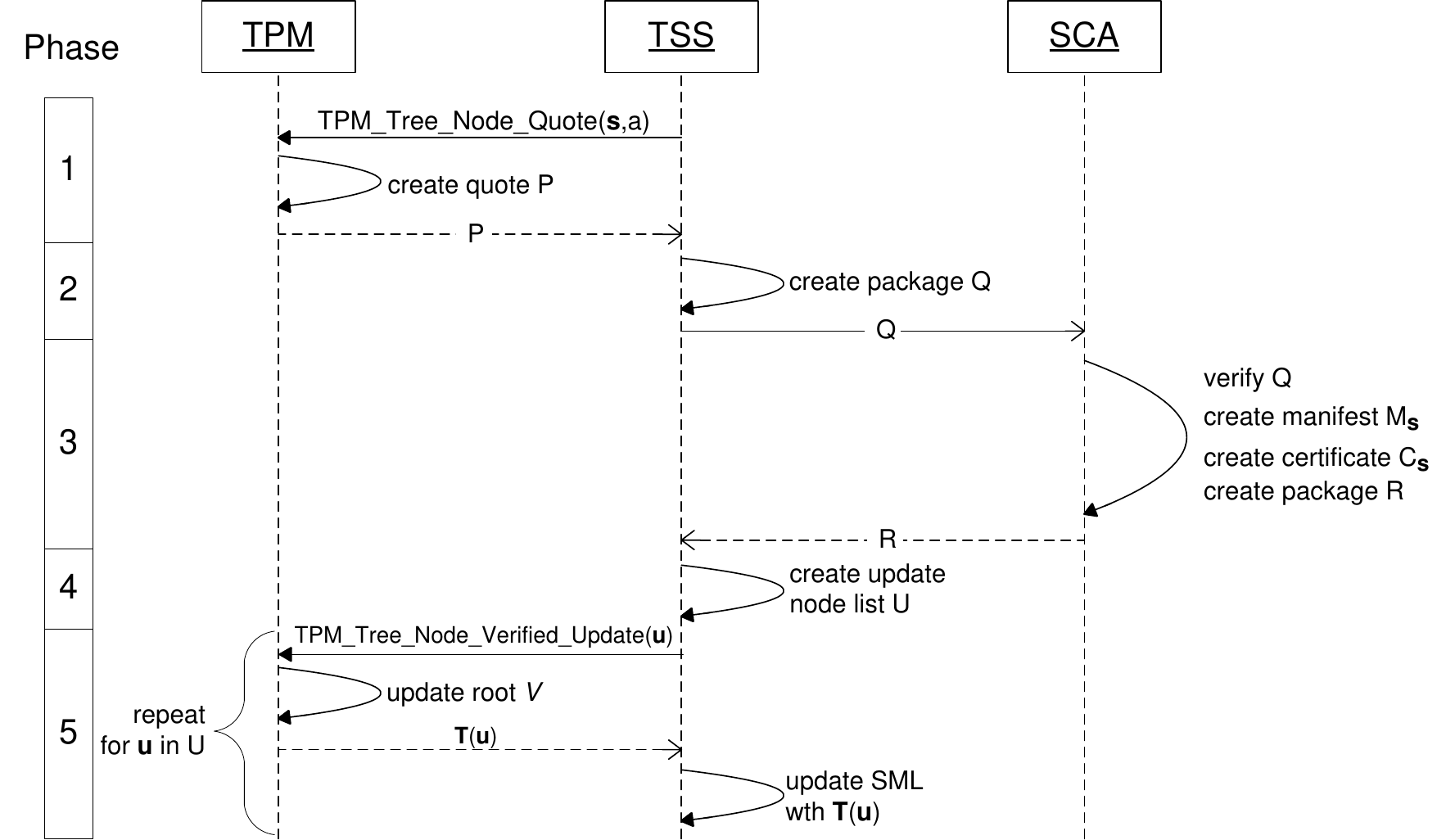}}
\caption{Certification protocol for a subtree with root $\mathbf{s}$.}
\label{fig:subtree-certification}
\end{figure*}
\section{Subtree Certification}\label{sec:subtr-cert}
We come to a primary use case of the command extensions introduced in Section~\ref{sec:tree-update}.
One of the biggest open problems of Trusted Computing is the association of semantics
to platform attestation. Existing TCG specifications define a bi-lateral remote attestation 
in which executed code is measured when it gets loaded. 
The measurements are stored in PCRs as verification data, 
and the TPM attests to these data by signing them with a TPM protected Attestation Identity Key (AIK).
Since a digest of a complete configuration is transmitted, 
the verifier needs to know all configurations of all machines (at all times, if system dynamics are 
considered).
The transmitted data for validation thus lacks expressiveness to enable versatile and efficient
remote platform validation. The need for semantic attestation was recognised early on
in~\cite{1267245} and later in~\cite{Baiardi2009}, 
who propose to restrict the scope of a single attestation to a virtualised
subsystem with limited complexity, allowing for attestation of complex, dynamic, and high-level program properties.  
In~\cite{Property-Attestation_IBM_2004} and~\cite{Sadeghi2004, Chen2006}  ``property,'' respectively, 
``property-based attestation'' (PBA) is proposed.
PBA allows to assure the verifier of security properties of the verified 
platform via a trusted third party (TTP), called \textit{Subtree CA} (SCA). 
The SCA issues a certificate which maps the platforms configuration 
to the properties (in particular desired/undesired functionality) which can be fulfilled in this configuration.
Essentially, PBA moves the infrastructural problem
of platform validation to a SCA, similarly to, but extending the role of, the TCG's privacy CA.
Certification of subtrees is one way to fill the mentioned ideas with life.
A related idea is that of hardware-supported updates~\cite{DBLP:conf/ches/KuhnKLSS05}, 
performed by a new, proposed TPM command, which re-seals data for another platform 
configuration based on an update certificate.
Also, in~\cite{sarmenta2006virtual} certificates over TPM-protected roots of hash tree are generated by a specific
TPM command, to certify the state of certain protected counter objects. Differently from that,
the present paper proposes a specific client-server protocol to obtain such a certificate from
a trusted thrid party.

Let us fix some notions.
In distinction to verification data, we call \textit{validation data} all data that can be submitted
to another party, the \textit{validator}, and used to assess the trustworthiness of the state
of the platform. The process of submission of validation data to the validator, 
for instance realised as remote attestation according to TCG,
and evaluation thereof by the validator, is properly called \textit{validation}.
Validation data may often comprise verification data such as quoted verification data
register (e.g., PCR) values. Validation may, beyond cryptographic verification of
verification data, include policy evaluation and triggering of actions by the validator.
See~\cite{AUS09B} for further discussion.

Tree-formed verification data and validation data associated
to an SML tree, provide structured data which can be used complementary to
the approaches above, to enhance the semantics of platform attestation.
Here we present one fundamental method using tree-formed verification data to realise concepts
related to PBA. Namely, it is shown how a SCA can replace a node in an SML tree with a meaningful,
trusted statement --- a \textit{subtree certificate} --- about the subtree of measured
components of which latter node is the root. This process realises a \textit{partial} validation
of the platform by the SCA and results in a trusted assertion that is ingested in the available
validation data of the platform. This can later be used toward another validator to validate the platform more fully.
In the next subsection we describe a generic base protocol for subtree certification which is common for
all conceivable variant realisations and use cases. After that, particularities and variants 
are considered in Subsections~\ref{sec:cert-subtr-bind}--~\ref{sec:subtree-discovery}.
\subsection{Subtree Certification Protocol}\label{sec:subtr-cert-prot}
Subtree certification is a process by which a Trusted Platform (the combination of
TPM and TSS in our simplified model) obtains a certificate for
the value of an inner node of a tree-formed SML from a SCA. 
For this, the platform submits a signed statement to the SCA, signifying that
the node value is contained in an SML tree of with root value protected in a TPM verification data
register. Based on this evidence , the SCA can issue a certificate with additional attributes
about this node value to the platform, which is then ingested into the SML tree.
The process essentially uses the protected tree operations introduced in Section~\ref{sec:tree-update}
above.

We assume that the platform possesses an active AIK $a$, with certificate $C_a$ issued by
a trusted Privacy CA (PCA).  
We further assume that communication between the platform and the SCA is encrypted and 
integrity-protected to mitigate man-in-the-middle attacks.
An inner node $\mathbf{s}$ is selected for certification. How this is done, shall concern
us later in Subsection~\ref{sec:subtree-discovery}.
In the protocol, we do not mention failure conditions and negative responses.
With these preparations, the subtree certification protocol proceeds in five phases, as
depicted in Figure~\ref{fig:subtree-certification}.

Phase 1 creates a quote over $\mathbf{s}$. For this, TSS calls 
\TQ with arguments $(\binr{\mathbf{s}},\mathbf{s},\mathbf{R}(\mathbf{s}),a)$
(note that Figure~\ref{fig:subtree-certification} shows only essential arguments
for brevity) and receives back
\[
P=\text{Sig}_a(\mathbf{s}).
\]
If the root of the tree, i.e. the register $V$ is selected for certification, then
$\texttt{TPM\_Quote}$ is to be used on $V$ instead.

In phase 2, the TSS creates an attestation package $Q$. 
It contains all necessary information for the verifying SCA, at least
\[
Q\subseteq\{P, \mathbf{s}, C_a, a_\text{pub}\}.
\]
(When the public part $a_\text{pub}$ of $a$ is not evident from $C_a$.
Also, the value of $\mathbf{s}$ may be known to SCA and then be omitted from $Q$). 
More information may be included as necessary, for instance the node coordinate
\binr{\mathbf{s}}, when it is part of the quote.
$Q$ is sent (encrypted with a public encryption key of SCA 
and integrity-protected) to SCA.
This phase is similar as in remote attestation specified by the TCG.

Phase 3 comprises the activities of SCA. First, SCA verifies $Q$ by verifying the signature of $P$
and tracing the certificate chain, up to the root certificate of the PCA, if necessary.
If the SCA recognises $\mathbf{s}$ as a node value which it can certify, it creates a manifest
$M_\mathbf{s}$ for it. This manifest may contain additional information about the platform state
associated with the presence of the subtree with root $\mathbf{s}$ in the SML, such as a time stamp,
a functionality of the platform, the identification of a module combined from the loaded components represented
by the leaf measurements of the subtree, or another platform property.
The manifest is the validation data added by subtree certification which provides
semantic meaning to the node value $\mathbf{s}$ to a validator.
Now, SCA can create a certificate for $\mathbf{s}$. This certificate, 
$C_\mathbf{s}$, binds the properties represented by $M$ to the platform, by binding it to the AIK $a$.
This can be done essentially in two ways, namely
\begin{equation}
  \label{eq:certsig}
  C_\mathbf{s} = 
\begin{cases}
  \text{Sig}_\text{SCA}(M_\mathbf{s}\| P) & \text{if $\mathbf{s}$ is revealed;}\\
  \text{Sig}_\text{SCA}(M_\mathbf{s}\| \text{bind}(a)) & \text{if $\mathbf{s}$ is concealed.}
\end{cases}  
\end{equation}
In the first case, SCA signs the manifest and the AIK-signed node value, thus establishing an
indirect binding to $a$. 
The binding of $C_\mathbf{s}$ to $a$ can then be verified if the platform reveals 
the node value $\mathbf{s}$.
In the second option, the binding is achieved directly, 
by letting the SCA sign some data $\text{bind}(a)$ which uniquely identifies $a$,
such as $a$'s public part, $C_a$, the serial number, or the fingerprint of $C_a$.
In the semantics of Public Key Infrastructures, $C_\mathbf{s}$ 
is, by the binding, an attribute certificate associated with $C_a$.
Finally, SCA creates a package $R$ containing at least $M_\mathbf{s}$ and $C_\mathbf{s}$, and
$\text{bind}(a)$ in the second case, and returns it to the platform.

Phase 4 prepares the update of the SML with certain data derived from $R$.
The SML update is an essential step to produce a binding association between
the subtree certificate and the certified node's position $\binr{\mathbf{s}}$ in the tree.
Only this allows the platform to assert to a validator that the property attested
by $C_\mathbf{s}$ and $M_\mathbf{s}$ is present in the platform's configuration.
Various ways of SML update to bind $C_\mathbf{s}$ to the represented subtree
are conceivable, each suited differently for particular use cases.
This is discussed in Subsection~\ref{sec:cert-subtr-bind}, while we now
state generic features of the SML update process.

A set $U=\{\mathbf{u}_1,\ldots,\mathbf{u}_k\}$ of new node nodes (values and positions in the
SML tree) is created with the following properties.  First, it
must hold $U\leq\mathbf{s}$, so that only the subtree below $\mathbf{s}$ is touched
by the update. This is necessary, since all
old SML tree nodes $\mathbf{n}\leq U$ strictly below $U$, i.e., $\mathbf{n}\not\in U$  are invalidated by the update,
and can not be verified anymore with respect to the updated root verification data register.
Second, $U$ is \textit{dependency-free}, i.e., 
\[
\nexists\ \mathbf{u},\ \mathbf{u}' \in U \colon \mathbf{u}\leq\mathbf{u}'.
\]
Dependency-freeness is the essential property ensuring consistency of the tree update
by $U$ with the one-way (upward) information flow embodied in Merkle hash trees. 
In particular it makes the update result independent of the order in which elements of $U$
are processed.

Phase 5 is the SML tree update proper. Iterating over  $\mathbf{u}\in U$, \vvupd is called with arguments
$(\binr{\mathbf{u}}, \mathbf{n}, \mathbf{u}, \mathbf{R}(\mathbf{n}))$,
where $\mathbf{n}$ is the old SML node value at position \binr{\mathbf{u}}.
This returns the new trace $\mathbf{T}(\mathbf{u})$ with which the TSS updates the SML.
Executing the tree update in the way described above maintains a consistent security level
for the SML and root verification data register. Namely, the operation \ex is always
executed inside the TPM.
When $U$ contains many elements, it may not be efficient to perform the update
in the way described for Phase 5, since \vvupd would in such a case verify many
overlapping reduced trees and thus incur redundancy in (complex) hash calculations.
A more efficient update algorithm is described in Appendix~\ref{sec:bulk update}.

A variant of the subtree certification protocol could combine the roles of PCA and SCA
for AIK, respectively, subtree certification in a single protocol run. An advantage would be that
no explicit generation and verification of an AIK certificate $C_a$ is necessary, because
generation, activation, and use of the AIK are bound into one session.
This combination of protocols is straightforward and left as an exercise to the reader.
\subsection{Certificate--Subtree Binding}\label{sec:cert-subtr-bind}
Binding the received subtree certificate to the platform state means binding it to
the tree-formed SML in the correct configuration, i.e., the position
of the certified subtree's root. As mentioned above, this is essential for meaningful 
subtree certificate-based validation in the context of an overall platform configuration.
One particular goal of binding $C_s$ to the SML tree is
integrity protection, since, for instance, later replacement with a different 
certificate must be prevented.
The binding can be achieved by updating parts of the tree with data which uniquely 
and verifiably identifies the subtree certificate. A wide range of data items can be produced from
the subtree certificate and entered into the SML tree in various positions.
Here, we describe some of the more sensible options.

In the simplest case the SML update may be trivial and $U$ may be empty.
This is only possible if $C_\mathbf{s}$ is composed by the first option of~(\ref{eq:certsig}),
 revealing $\mathbf{s}$. Then $\mathbf{s}$ can just be retained in the SML tree and whether
the subtree below it is also retained depends on the use case.
The binding association is via the actual node value $s$ signed by $C_\mathbf{s}$.

As another example, consider the case that all meaningful data concerning the platform property
attested by the certificate should be protected by the updated tree, e.g., for forensic use.
That is, the three data items $\mathbf{s}$, $M_\mathbf{s}$, and $C_\mathbf{s}$ shall enter the
update set. While the node value $\mathbf{s}$ is already in the correct data format, the other
two are first processed to $m(M_\mathbf{s})$ and $m(C_\mathbf{s})$. The operation $m$ can be the
generation of a hash value by the platform's Root of Trust for Measurement (RTM), or another
appropriate one-way operation. If some data item already contains suitable, uniquely identifying
data of the appropriate node value format, then it can be directly extracted and used as node
update value. A particular example could be a certificate fingerprint contained in $C_\mathbf{s}$.
The three update nodes can then be configured in an update set to produce, for instance,
the following configuration of updated nodes in the SML tree.
\begin{center}
\begin{tikzpicture}[level distance=6mm,level/.style={sibling distance=24mm/#1}]
\node [style={rectangle}] {$( k , \binr{\mathbf{s}})$}
child { 
  node [style={rectangle, inner sep=3.5pt}] {$\ast$}
  child {
    node [style={rectangle}] {$m(C_\mathbf{s})$} 
  }
  child {
    node [style={rectangle}] {$m(M_\mathbf{s})$} 
  }
}
child { 
  node [style={rectangle, inner sep=3.5pt}] {$\mathbf{s}$} 
}; 
\end{tikzpicture}
\end{center} 
The root of the updated subtree is inserted in the old position of $\mathbf{s}$ and 
has the value $k=(m(C_\mathbf{s})\ex m(M_\mathbf{s}) )\ex \mathbf{s}$.
This configuration provides independent integrity protection to the subtree
certificate and manifest, and retains the old node value independently.
In particular, attestation to the platform property represented by $C_\mathbf{s}$ can, in this configuration,
still be done without revealing information about $\mathbf{s}$, by quoting only the left inner
node $\ast$ of the subtree.

Variants of certificate to subtree binding abound. The platform may also want to 
include (integrity protection values of) own generated data therein, for instance 
an internal time stamp from a secure clock.
What makes sense depends ultimately on the use case.
\subsection{Subtree Validation}\label{sec:subtree-validation}
For the attestation of the property represented by a subtree certificate to a validator,
the platform can quote, using \TQ, any node in or above the updated subtree which protects the intended
validation data, which comprises at least the manifest and the certificate proper.
The platform will then submit validation data as necessary to the validator, at least
all data needed for verification of the asserted property, again comprising at least
$M_\mathbf{s}$ and $C_\mathbf{s}$. Note that the validation data which is already
protected by the submitted quote does in principle not require additional integrity protection
in this.

One important point for the validator is to verify platform binding of the validation data.
It is known that proving this property, i.e., that the validating platform is the same
that performed subtree certification toward the SCA, is non-trivial~\cite{AndreasFuchs2009}.
The simplest way to achieve it is to use the \textit{same} AIK, $a$, in subtree validation as in
certification. The platform would then also submit $a_\text{pub}$, and if necessary also $C_a$
as part of the validation data. Whether $C_a$ is needed depends on the semantics of the
subtree certificate, i.e., SC may already have checked the AIK certificate and $C_\mathbf{s}$ may 
state its veracity. According information can be placed in the manifest.
Re-using the same AIK partially compromises privacy, and other methods to solve the
problem may be worth further study.
\subsection{Subtree Discovery}\label{sec:subtree-discovery}
An important step for the practical use of subtree certification is the discovery of
subtrees for which a platform can obtain certificates from a particular SCA.
Without going into details of the interactions between platform, SCA, and validator,
two categories of subtree discovery methods are described below.
The first one places the workload of subtree discovery with the platform, while the second
one assumes a ``dumb'' platform and places more load on the SCA.
\subsubsection{Active Discovery}
In this model, the SCA sends some \textit{subtree discovery data} to the platform, in the simplest case
a list of node values which it is ready to certify.
The platform can search for these values in its SML tree and perform subtree certification for
each identified node.
This baseline procedure suggests various refinements, in particular enriching the discovery data
and extending the discovery procedure to a negotiation protocol between platform and SCA.
For instance, discovery data may contain root node positions as conditions on certifiable roots,
which would, in the case of an SML produced in an authenticated boot process, correspond to the
fact that the components loaded during the build of the latter subtree are loaded at a defined 
stage of the platform start up. Such conditions on absolute positioning of nodes may be difficult
in practise for complex platforms whose configurations may change dynamically.
More refined conditions could therefore also express relative positions of some, e.g., pairs of
certifiable roots. The SC could state expressions saying ``$\mathbf{s}$ is certifiable, if
it is preceded by $\mathbf{r}$'' (i.e., $\mathbf{r}$ lies to the left of $\mathbf{s}$ in the
ordered SML tree). This can be interpreted to the end that a certain functionality is
operational on the platform only if another functionality was made operational before it.

A more fundamentally different variant of Model~I is that the discovery data does not consist
of subtree roots, i.e., inner nodes, but rather of leaf, i.e., measurement, values.
A ``bottom up'' discovery procedure would require that the platform makes an ``educated guess''
about which inner nodes are certifiable, based on the received leaf measurement values
which the SCA asserts to know as trusted values. A simplistic method is to
find the set of span roots of subtrees whose leaves are all in the discovery data.
The platform may then quote a subtree root and send it together with its SML subtree
to the SCA. In general, the SCA will have to verify the SML subtree and decide if it
is ready to certify that root, since this may still depend on the order of the leaves.
In many cases, the platform may want to obtain a certificate for a subtree for which
the particular SCA knows only \textit{some} leaf values, i.e., the leaf set of the corresponding 
subtree has gaps with respect to the discovery data.
If the platform has other trusted data, for instance RIM certificates~\cite{MTMREFARC} obtained from
a party which the SCA trusts, the platform could submit these data in Phase~2 of the subtree certification,
to aid SCA with its decision to certify the subtree root.
\subsubsection{Passive Discovery}
In the case that the device is not capable to perform a local discovery of 
subtrees, a model can be used which moves the computations to the SCA.
The platform selects an inner node $\mathbf{n}$, with the aim to retrieve certificates
from the SCA for any suitable subtrees below $\mathbf{n}$.
The node $\mathbf{n}$ could be equal to the root of the complete tree ($V$), 
in the case that the platform wants to get all certifiable nodes certified.
The next two steps are the same as described in Phases~1 and~2 in section~\ref{sec:subtr-cert-prot},
i.e., the platform performs a \TQ, or $\texttt{TPM\_Quote}$, respectively, if $V$ was selected.
Wrapped in an attestation package together with the SML subtree below the quoted root.
The SCA receives this information and can then,
using tree traversal techniques described in~\cite{tree01},
verify the integrity of the tree and concurrently 
find one or multiple (disjoint) subtrees $S_i$ with certifiable set of roots $S$.
The SCA then iterates phase 3 of the protocol from~\ref{sec:subtr-cert-prot}, creating certificates
for all $\mathbf{s}_i\in S$.
Since the protocol allows for the update of multiple nodes, incorporated into the
update node list $U$, the update of all found subtrees can be done in a single protocol run.
A variant could be for the platform to not send a \TQ or $\texttt{TPM\_Quote}$
in the first step, but only provide the SCA with the SML, starting from the selected node $\mathbf{n}$.
The SCA then searches for potential candidate subtrees to be certified and then requests
the platform to provide a \TQ for the root nodes of the identified subtrees.
This is a trade-off in the sense that it allows the platform to send the SML without
 having to perform cryptographic operations in advance.
Nevertheless, the platform must provide appropriate quotes prior to the certification
by the SCA to provide integrity protection for the sent SML.

\section{Conclusion}\label{sec:conclusion}
%
%
Validation using tree-formed SMLs and verification data registers adds semantics
to remote attestation. The possibility to attest to subtrees of an SML enables 
expressiveness far beyond conventional remote attestation. Tree-formed verification
data is a promising way to substantiate other proposals to add semantics to platform 
attestation, e.g., association and of properties to validation data,  as in PBA. 

Further work shall pursue this promising direction and consider concrete architectures
for platform validation with tree-formed verification and validation data --- 
what we call \textit{tree-formed validation} (TFV).
One conceivable option is to efficiently organise a database
of reference trees by an SCA and/or a validator in a way that allows for the 
modular building using subtrees of known component sub-structures, e.g., 
dependent programs loaded in sequence, or
components with uniform associated security policies. 
Architectures and methods for subtree discovery, expression of dependencies
between validated platform components, and management (updates, remediation)
of platforms according to results of TFV are subjects
of ongoing research and shall be discussed elsewhere.
\appendices
\section{Efficient, Secure Node Set Update}\label{sec:bulk update}
As mentioned in Subsection~\ref{sec:subtr-cert-prot}, the efficiency of updating a large set
$U$ of inner nodes using \vvupd depends on the overlap of the reduced trees of the
elements of $U$, since many redundant hash
operations for verification can occur.
A bulk update strategy can be applied to improve the na\"ive algorithm
of Phase 5 of the subtree certification protocol, using only the 
\TExt command described in~\cite{tree01}.
It rests on the observation that subsets of the update set $U$ span subtrees which
are independent of the old SML values, i.e., their roots depend only on nodes in $U$.
Thus the roots of the trees spanned by such sets can be pre-calculated without
expensive verification.

We first need some definitions.
Assume $U=\{\mathbf{u}_1,\ldots,\mathbf{u}_k\}$ is a dependency-free update set.
A node in the SML tree is called $U$\textit{-intrinsic}, if a) it is an
element of $U$, b) its sibling is in $U$, or c) its sibling is $U$-intrinsic.
This recursive definition captures all nodes whose updated values depend only 
on $U$ and not on SML nodes in the complement of $U$.
The \textit{span root} of a subset $V\subseteq U$ is the unique intersection point of the
traces of all elements of $V$.
The subtree \textit{spanned} by a subset $V\subseteq U$ is the union of all traces of
elements of $V$ with all nodes strictly above the span root omitted.
Now, the subset $V$ is called $U$\textit{-intrinsic} iff all elements of its
spanned subtree are $U$-intrinsic.

With these settings, more efficient update of the SML with $U$ is done as follows.
\begin{enumerate}
\item Identify the (mutually disjoint) $U$-intrinsic subsets $V_1,\ldots,V_k\subseteq U$.
\item Iterate over $V_i$, $i=1,\ldots,k$.
  \begin{enumerate}
  \item Normalise the coordinates of elements of $V_i$ by
    \begin{enumerate}
    \item truncating the prefix given by the coordinate of the span root of $V_i$, and
    \item post-fixing zeroes until all coordinates have equal length, the depth of the
          subtree spanned by $V_i$.
    \end{enumerate}
  \item Order the elements of $V_i$ alphabetically according to their normalised coordinates, producing
an ordered list $\widetilde{V}_i$.
  \item Fill up all gaps (in the normalised coordinates) in  $\widetilde{V}_i$ with \nil values.
  \item Select a free verification data register $V'$.
  \item Sequentially use \TExt on the elements of $\widetilde{V}_i$ with target $V'$.
  \item Remove $V_i$ from $U$.
  \item Insert $(V',\binr{\mathbf{v}_i})$ into $U$, where $\mathbf{v}_i$ is $V_i$'s
        span root.
  \end{enumerate}
\item For the remaining elements of $U$, apply the normal update procedure of Phase 5
in Subsection~\ref{sec:subtr-cert-prot} using
\vvupd.
\end{enumerate}
%
%
\section*{Acknowledgement}
%
%
This work was funded by InterDigital, Inc. Special thanks go to
Lawrence Case, 
Bob DiFazio,
David Greiner,
Louis Guccione, 
Dolores Howry,
Michael V. Meyerstein,
and
Sudhir Pattar, 
for many useful discussions and comments.




\begin{thebibliography}{10}
\providecommand{\url}[1]{#1}
\csname url@samestyle\endcsname
\providecommand{\newblock}{\relax}
\providecommand{\bibinfo}[2]{#2}
\providecommand{\BIBentrySTDinterwordspacing}{\spaceskip=0pt\relax}
\providecommand{\BIBentryALTinterwordstretchfactor}{4}
\providecommand{\BIBentryALTinterwordspacing}{\spaceskip=\fontdimen2\font plus
\BIBentryALTinterwordstretchfactor\fontdimen3\font minus
  \fontdimen4\font\relax}
\providecommand{\BIBforeignlanguage}[2]{{%
\expandafter\ifx\csname l@#1\endcsname\relax
\typeout{** WARNING: IEEEtran.bst: No hyphenation pattern has been}%
\typeout{** loaded for the language `#1'. Using the pattern for}%
\typeout{** the default language instead.}%
\else
\language=\csname l@#1\endcsname
\fi
#2}}
\providecommand{\BIBdecl}{\relax}
\BIBdecl

\bibitem{AUS09B}
A.~U. Schmidt, I.~Cha, and A.~Leicher, ``Scaling concepts between trust and
  enforcement,'' in \emph{Trust Modeling and Management in Digital
  Environments: From Social Concept to System Development}, Z.~Yan, Ed.\hskip
  1em plus 0.5em minus 0.4em\relax IGI Global Publishing, 2010, pp. 20--57.

\bibitem{PCCLIENTBIOS}
{T}rusted~{C}omputing {G}roup, ``{TCG PC Client Specific Implementation
  Specification For Conventional BIOS},'' Version 1.20 FINAL Revision 1.00,
  July 2005, for TPM Family 1.2; Level 2.

\bibitem{MTMREFARC}
\BIBentryALTinterwordspacing
------, ``{TCG Mobile Reference Architecture},'' Specification version 1.0
  Revision 5, TCG, June 2008. [Online]. Available:
  \url{https://www.trustedcomputinggroup.org/}
\BIBentrySTDinterwordspacing

 \bibitem{tree01} A.\ Schmidt and A.\ Leicher, ``Tree-formed Verification Data for Trusted Platforms,''
 preprint, InterDigital, Inc., Novalyst IT, November 2009. [Online]. Available:
  \url{http://arxiv.org/abs/1007.0642}

\bibitem{TPMMAIN}
\BIBentryALTinterwordspacing
------, ``{TPM} {M}ain,'' Specification Version 1.2 Level 2 Revision 103, TCG,
  July 2007. [Online]. Available: \url{https://www.trustedcomputinggroup.org}
\BIBentrySTDinterwordspacing

\bibitem{Merkle1989}
R.~C. Merkle, ``A certified digital signature,'' in \emph{Advances in
  Cryptology (CRYPTO '89)}, ser. LNCS, G.~Brassard, Ed., no. 435.\hskip 1em
  plus 0.5em minus 0.4em\relax Springer-Verlag, 1989, pp. 218--238,
  republication of the 1979 original.

\bibitem{10.1109/SP.1980.10006}
------, ``Protocols for public key cryptosystems,'' in \emph{Ptroceedings of
  the IEEE Symposium on Security and Privacy}, vol.~0.\hskip 1em plus 0.5em
  minus 0.4em\relax Los Alamitos, CA, USA: IEEE Computer Society, 1980, p. 122.

\bibitem{DBLP:conf/acsac/MoyerBSMJ09}
T.~Moyer, K.~R.~B. Butler, J.~Schiffman, P.~D. McDaniel, and T.~Jaeger,
  ``Scalable web content attestation,'' in \emph{Proceedings of the
  Twenty-Fifth Annual Computer Security Applications Conference, ACSAC 2009,
  Honolulu, Hawaii, 7-11 December 2009}.\hskip 1em plus 0.5em minus 0.4em\relax
  IEEE Computer Society, 2009, pp. 95--104.

\bibitem{1267245}
\BIBentryALTinterwordspacing
V.~Haldar, D.~Chandra, and M.~Franz, ``Semantic remote attestation: a virtual
  machine directed approach to trusted computing,'' in \emph{VM'04: Proceedings
  of the 3rd conference on Virtual Machine Research And Technology
  Symposium}.\hskip 1em plus 0.5em minus 0.4em\relax Berkeley, CA, USA: USENIX
  Association, 2004, pp. 29--41. [Online]. Available:
  \url{http://www.usenix.org/events/vm04/tech/haldar.html}
\BIBentrySTDinterwordspacing

\bibitem{Baiardi2009}
F.~Baiardi, D.~Cilea, D.~Sgandurra, and F.~Ceccarelli, ``{Measuring Semantic
  Integrity for Remote Attestation},'' in \emph{Proceedings of the 2nd
  International Conference on Trusted Computing}.\hskip 1em plus 0.5em minus
  0.4em\relax Springer, 2009, pp. 81--100.

\bibitem{Property-Attestation_IBM_2004}
J.~Poritz, M.~Schunter, E.~V. Herreweghen, and M.~Waidner, ``{Property
  Attestation--Scalable and Privacy-friendly Security Assessment of Peer
  Computers},'' IBM Zurich Research Laboratory, R{\"u}schlikon, Switzerland,
  Research Report RZ 3548, 10 2004.

\bibitem{Sadeghi2004}
A.~Sadeghi and C.~St{\"u}ble, ``{Property-based attestation for computing
  platforms: caring about properties, not mechanisms},'' in \emph{Proceedings
  of the 2004 workshop on New security paradigms}.\hskip 1em plus 0.5em minus
  0.4em\relax ACM New York, NY, USA, 2004, pp. 67--77.

\bibitem{Chen2006}
L.~Chen, R.~Landfermann, H.~L{\"o}hr, M.~Rohe, A.~Sadeghi, and C.~St{\"u}ble,
  ``{A protocol for property-based attestation},'' in \emph{Proceedings of the
  first ACM workshop on Scalable trusted computing}.\hskip 1em plus 0.5em minus
  0.4em\relax ACM New York, NY, USA, 2006, pp. 7--16.

\bibitem{DBLP:conf/ches/KuhnKLSS05}
U.~K{\"u}hn, K.~Kursawe, S.~Lucks, A.-R. Sadeghi, and C.~St{\"u}ble, ``Secure
  data management in trusted computing,'' in \emph{Proceedings of the 7th
  International Workshop on Cryptographic Hardware and Embedded Systems - CHES
  2005, Edinburgh, UK, August 29 - September 1, 2005,}, ser. Lecture Notes in
  Computer Science, J.~R. Rao and B.~Sunar, Eds., vol. 3659.\hskip 1em plus
  0.5em minus 0.4em\relax Springer, 2005, pp. 324--338.

\bibitem{sarmenta2006virtual}
L.~Sarmenta, M.~van Dijk, C.~O'Donnell, J.~Rhodes, and S.~Devadas, ``{Virtual
  monotonic counters and count-limited objects using a TPM without a trusted
  OS},'' in \emph{Proceedings of the first ACM workshop on scalable trusted
  computing}.\hskip 1em plus 0.5em minus 0.4em\relax ACM, 2006, p.~42.

\bibitem{AndreasFuchs2009}
A.~Fuchs, S.~G{\"u}rgens, and C.~Rudolph, ``On the security validation of
  integrated security solutions,'' in \emph{Emerging Challenges for Security,
  Privacy and Trust, 24th IFIP TC 11 International Information Security
  Conference, SEC 2009, Pafos, Cyprus, May 18-20, 2009. Proceedings}, ser. IFIP
  Advances in Information and Communication Technology, D.~Gritzalis and
  J.~L{\'o}pez, Eds., vol. 297.\hskip 1em plus 0.5em minus 0.4em\relax Boston,
  MA, USA: Springer, 2009, pp. 190--201.

 \end{thebibliography}
%

\providecommand{\noopsort}[1]{} \providecommand{\singleletter}[1]{#1}

\end{document}